 \documentclass[a4paper]{article}
 \usepackage{etoolbox}
 \newbool{PREPRINT}
 \booltrue{PREPRINT}
\usepackage{amssymb,amsmath,listings, algorithm, algcompatible, subcaption, graphicx}
\usepackage[colorlinks,bookmarksopen,bookmarksnumbered,citecolor=red,urlcolor=red]{hyperref}
\usepackage[usenames]{color}
\usepackage{xcolor}

\renewcommand{\vec}[1]{\boldsymbol{#1}}

\newcommand{\sr}{\ensuremath{\text{sr}}}
\newcommand{\lr}{\ensuremath{\text{lr}}}

\definecolor{DarkBlue}{rgb}{0.00,0.00,0.55}
\definecolor{DarkRed}{rgb}{0.55,0.00,0.00}
\definecolor{DarkGreen}{rgb}{0.00,0.55,0.00}
\definecolor{Gray}{rgb}{0.95,0.95,0.95}
\definecolor{Purple}{rgb}{0.5,0.0,0.5}
\definecolor{Bittersweet}{rgb}{1.0,0.44,0.37}

\lstdefinelanguage[ppmd]{python}[]{python}{%
  emph={ParticleLoop,ParticleDat,PositionDat,ScalarArray,Kernel,PairLoop,Constant,State,IntegratorRange}
}

\ifbool{PREPRINT}{ % PREPRINT
\usepackage[colorlinks,bookmarksopen,bookmarksnumbered,citecolor=red,urlcolor=red]{hyperref}
\usepackage[affil-it]{authblk}
\usepackage{fullpage}

}{ % PREPRINT
\begin{document}
\def\hb{\hbox to 10.7 cm{}}
\pagestyle{headings}
\def\thepage{}

\begin{frontmatter}              % The preamble begins here.
}

\title{Long range forces in a performance portable Molecular Dynamics framework
\ifbool{PREPRINT}{\thanks{Submitted to ParCo2017}}{}
}
\ifbool{PREPRINT}{ % PREPRINT
\author[a,$\dagger$]{William~Robert~Saunders}
\author[b]{James~Grant}
\author[a]{Eike~Hermann~M\"{u}ller}
\affil[ ]{University of Bath, Bath BA2 7AY, United Kingdom}
\affil[a]{Department of Mathematical Sciences}
\affil[b]{Department of Chemistry}
\affil[$\dagger$]{Email: \texttt{w.r.saunders@bath.ac.uk}}
\begin{document}
\maketitle
}{ % PREPRINT
\markboth{}{September 2017\hb}
\author[A]{\fnms{William Robert} \snm{Saunders}%
\thanks{Corresponding Author: E-mail:
\texttt{w.r.saunders@bath.ac.uk}}},
\author[B]{\fnms{James} \snm{Grant}}
and
\author[A]{\fnms{Eike Hermann} \snm{M\"{u}ller}}
\runningauthor{W.R. Saunders et al.}
\address[A]{Department of Mathematical Sciences}
\address[B]{Department of Chemistry}
\address{University of Bath, Bath BA2 7AY, United Kingdom}
}
\begin{abstract}
Molecular Dynamics (MD) codes predict the fundamental properties of matter by following the trajectories of a collection of interacting model particles. To exploit diverse modern manycore hardware, efficient codes must use all available parallelism. At the same time they need to be portable and easily extendible by the domain specialist (physicist/chemist) without detailed knowledge of this hardware. To address this challenge, we recently described a new Domain Specific Language (DSL) for the development of performance portable MD codes based on a ``Separation of Concerns'': a Python framework automatically generates efficient parallel code for a range of target architectures.

Electrostatic interactions between charged particles are important in many physical systems and often dominate the runtime. Here we discuss the inclusion of long-range interaction algorithms in our code generation framework. These algorithms require global communications and careful consideration has to be given to any impact on parallel scalability. We implemented an Ewald summation algorithm for electrostatic forces, present scaling comparisons for different system sizes and compare to the performance of existing codes. We also report on further performance optimisations delivered with OpenMP shared memory parallelism.
\end{abstract}

\ifbool{PREPRINT}{ % PREPRINT                                                   
% === ACM classifiers ===                                                       
\textbf{keywords}:
\newcommand{\sep}{, }
}{ % PREPRINT
\begin{keyword}
}
Molecular Dynamics\sep Electrostatic\sep Ewald Summation\sep Domain Specific Language\sep Parallel Computing
% === ACM classifiers ===
% G.4 MATHEMATICAL SOFTWARE -> Portability, Parallel & vector implementations
% D.1.3 PROGRAMMING TECHNIQUES Concurrent Programming -> Parallel Programming
% D.2.11 SOFTWARE ENGINEERING -> Software Architectures -> Domain-specific architectures
% J.2 PHYSICAL SCIENCES AND ENGINEERING -> Chemistry, Physics
\ifbool{PREPRINT}{ % PREPRINT
}{\end{keyword}
\end{frontmatter}
\markboth{September 2017\hb}{September 2017\hb}
}

%%%%%%%%%%%%%%%%%%%%%%%%%%%%%%%%%%%%%%%%%%%%%%%%%%%%%%%%%%%%%%%%%%%%%
\section{Introduction}
%%%%%%%%%%%%%%%%%%%%%%%%%%%%%%%%%%%%%%%%%%%%%%%%%%%%%%%%%%%%%%%%%%%%%
Molecular Dynamics (MD) codes are well established computational tools for predicting the behaviour of complex physical, chemical and biological systems. They can be used to replace expensive laboratory experiments or to perform simulations in experimentally inaccessible areas of parameter space. Classical MD codes model a material by following a large number $N$ of interacting particles, which obey Newton's laws of motion. Since statistical fluctuations are suppressed with inverse powers of $N$, to overcome finite size effects and to study increasingly complex systems, calculations have $N\gg 1$ and require substantial computational power.

To make efficient use of large HPC installations, developers of MD codes face several challenges when exploiting the hierarchical parallelism of modern manycore chip architectures. Unfortunately it is rare for a computational physicist or chemist to be an expert both in their domain and in the low level optimisation of parallel codes. This impedes the development of fast yet complex MD codes and limits scientific productivity. One solution to this problem, which has been successfully applied in other fields such as grid-based partial differential equation solvers \cite{Bertolli2012,Rathgeber2012}, is the introduction of a ``Separation of Concerns''. This allows the domain specialist to describe the problem at a high abstraction level independent of the hardware, while a computational scientist provides a mechanism for automatically executing this high level representation efficiently on various parallel architectures. 

In a previous paper \cite{ppmd2017} we described a domain specific language (DSL) for the implementation of performance portable MD codes. The key observation is that interactions between pairs of particles can be expressed as small kernels. While the domain specialist writes this kernel as a short piece of C-code, the hardware-dependent execution over all particle pairs is realised via a code generation system, which automatically produces efficient looping code. As demonstrated in \cite{ppmd2017}, the resulting performance both for a multi-CPU and multi-GPU implementation is comparable with that of well established monolithic MD codes such as LAMMPS \cite{Plimpton1995} and DL-POLY \cite{Todorov2006} for a short-range Lennard-Jones benchmark.

More generally, the potential between two particles can be split into a short- and long-range part. In many applications the computational bottleneck is the calculation of electrostatic forces between charged particles. Support for long range interactions has been missing so far in the framework described in \cite{ppmd2017}. In this paper we report on the implementation of a classical Ewald summation technique for the inclusion of electrostatic interactions within our framework and present parallel performance results for a system of interacting charged particles.

Parallel scalability can be improved by using shared memory parallelism within a node since this reduces communication costs and load imbalance. Here we also report on a new hybrid MPI+OpenMP backend for the framework in \cite{ppmd2017}.
%We present some first results on a hybrid MPI+OpenMP implementation which improves on the strong scaling compared with MPI-only results in \cite{ppmd2017}.
\ifbool{PREPRINT}{ % PREPRINT                                                  
\paragraph{Structure.} This paper is organised as follows: In Section \ref{sec:ppmd} we briefly review some key concepts of MD simulations with particular focus on long range force calculations; here we also summarise the design principles of the abstraction in \cite{ppmd2017}. The Ewald summation algorithm and its implementation in our code generation framework is described in Section \ref{sec:methods}. We present numerical results in Section \ref{sec:results} and conclude in Section \ref{sec:conclusion}.
}{}
%%%%%%%%%%%%%%%%%%%%%%%%%%%%%%%%%%%%%%%%%%%%%%%%%%%%%%%%%%%%%%%%%%%%%
\section{Molecular Dynamics in a performance portable framework}\label{sec:ppmd}
%%%%%%%%%%%%%%%%%%%%%%%%%%%%%%%%%%%%%%%%%%%%%%%%%%%%%%%%%%%%%%%%%%%%%
%%%%%%%%%%%%%%%%%%%%%%%%%%%%%%%%%%%%%%%%%%%%%%%%%%%%%%%%%%%%%%%%%%%%%
\subsection{Long range forces in Molecular Dynamics}
%%%%%%%%%%%%%%%%%%%%%%%%%%%%%%%%%%%%%%%%%%%%%%%%%%%%%%%%%%%%%%%%%%%%%
In a MD simulation the force on each particle has to be calculated at every timestep; this is often the bottleneck of the model run. For conservative forces, the force on particle $i$ can be written as the gradient of a phenomenological potential $U(\{\vec{r}_j\})$, which depends on the positions $\{\vec{r}_j\}$, $j=1,\dots,N$ of all particles. We only consider two-particle potentials and split $U$ into a short-range (\sr) part and long-range (\lr) contribution to obtain the total force on particle $i$ as
\begin{xalignat}{2}
\vec{F}_i &= -\frac{\partial}{\partial\vec{r}_i}U(\{\vec{r}_j\}),&
U(\{\vec{r}_j\}) = \sum_{j\ne i}\left(U^{(\sr)}(\vec{r}_i,\vec{r}_j) +U^{(\lr)}(\vec{r}_i,\vec{r}_j)\right).\label{eqn:force}
\end{xalignat}
The short-range potential $U^{(\sr)}$ (such as Lennard-Jones interactions) can be safely truncated beyond a distance $r_c$ and only a fixed number of neighbours $j$ need to be considered for a given particle $i$. As a result the total cost of the force calculation is $\mathcal{O}(N)$ (see e.g. \cite{rapaport_art}). This, however, is not true for electrostatic interactions for which the potential decays with the inverse separation of the particles
\begin{xalignat*}{2}
U^{(\lr)}(\vec{r}_i,\vec{r}_j) &= q_i\phi_j(|\vec{r}_i-\vec{r}_j|)&\text{with}\qquad
\phi_j(r) \equiv \frac{q_j}{r}.
\end{xalignat*}
To see this consider the contributions arising from a particle $i$ interacting with a uniform charge density, $\rho$ at distances greater than the cut-off $r_c$. If this contribution is neglected, the missing energy diverges as $U_{\text{exact}}-U_{\text{truncated}} \approx \frac{N}{2}\int_{r_c}^\infty \frac{\rho}{r} 4\pi r^2\;dr = 2\pi N\rho\int_{r_c}^\infty r\;dr$. As a consequence, computing the long range potential and the resulting force is naively an $\mathcal{O}(N^2)$ operation, since for each particle all $N-1$ neigbours have to be considered. In a parallel implementation this requires global communication of all particle positions and charges. Even worse, if periodic boundary conditions are used, another sum over all periodic images of the simulation domain is necessary and the resulting sum will not necessarily converge. However, for neutral systems algorithms exist for computing the electrostatic interaction by a suitable re-ordering of the sums over particle pairs. The computational complexity is reduced to $\mathcal{O}(N^{3/2})$ with a classic Ewald method \cite{Ewald1921}, as described in Section \ref{sec:methods}. This method is suitable for small to  medium-size systems and in this work we explain how it can be implemented in the performance portable framework described in \cite{ppmd2017}. Smooth Particle-Mesh Ewald (SPME) methods use the Fast Fourier transform to reduce this complexity further to $\mathcal{O}(N\log(N))$ \cite{spme2, spme3} and the Fast Multipole Method \cite{fmm1} can achieve optimal $\mathcal{O}(N)$ complexity; since their implementation is more challenging, those approaches will be considered in a subsequent study.
%%%%%%%%%%%%%%%%%%%%%%%%%%%%%%%%%%%%%%%%%%%%%%%%%%%%%%%%%%%%%%%%%%%%%
\subsection{Abstraction and Python code generation framework}
%%%%%%%%%%%%%%%%%%%%%%%%%%%%%%%%%%%%%%%%%%%%%%%%%%%%%%%%%%%%%%%%%%%%%
To calculate quantities such as the force in Eq. (\ref{eqn:force}) requires looping over all particle pairs in an MD simulation and we now describe how this is implemented in our code generation framework. The abstraction in \cite{ppmd2017} assumes that (i) the physical system can be described by assigning a set of properties (such as mass, charge, position, velocity) to each model particle and that (ii) all computationally expensive operations in the MD code can be realised by looping over all particles, or all pairs of particles. Those operations are encoded in a local pairwise kernel which only operates on the properties on the two participating particles. For example, the kernel might increment the total force on a particle by adding the force exerted by another particle. Loops over individual particles are implemented in a similar way. The DSL is realised as a Python code generation framework, which is available at \url{https://bitbucket.org/wrs20/ppmd}.
Individual particle properties such as position, velocity and charge are stored as Python \texttt{ParticleDat} objects; global properties shared by all particles such as the total potential energy are realised as \texttt{GlobalArray} objects. To describe a pairwise kernel, the user writes a short, hardware independent piece of C-code for the manipulation of the local properties, see Listing \ref{lst_sr} for an example. This is launched via a Python \texttt{PairLoop} call which specifies the accessed \texttt{ParticleDat}s and \texttt{GlobalArray}s together with access descriptors (see Listing \ref{lst_pysr}). The access descriptors are used to trigger suitable halo exchanges and reductions in \texttt{GlobalArray} objects.

We stress, however, that the user never has to write any explicit calls to MPI routines or add OpenMP directives; the parallelisation is implicit. When the user runs the Python code, a hardware dependent C-wrapper library which executes the kernel over all particle pairs is generated, compiled and run. This allows the execution of the loop at the speed of a compiled language while still allowing the user to express the overarching algorithms (such as the MD time stepping method) in a high-level Python framework. To increase efficiency the \texttt{PairLoop} call can be specialised for local interactions to obtain a \textit{local} \texttt{PairLoop}. For this only pairs of particles which are separated by less than a given cutoff distance $r_c$ are considered and efficient $\mathcal{O}(N)$ pair looping wrapper code is generated.
\ifbool{PREPRINT}{%

Before returning to the implementation of long range interactions in our framework in Section \ref{sec:implementation} we describe the Particle-Ewald summation technique in \cite{Ewald1921} for calculating electrostatic interactions.
}{}
%%%%%%%%%%%%%%%%%%%%%%%%%%%%%%%%%%%%%%%%%%%%%%%%%%%%%%%%%%%%%%%%%%%%%
\section{Methods}\label{sec:methods}
%%%%%%%%%%%%%%%%%%%%%%%%%%%%%%%%%%%%%%%%%%%%%%%%%%%%%%%%%%%%%%%%%%%%%
%%%%%%%%%%%%%%%%%%%%%%%%%%%%%%%%%%%%%%%%%%%%%%%%%%%%%%%%%%%%%%%%%%%%%
\subsection{Ewald Summation}
%%%%%%%%%%%%%%%%%%%%%%%%%%%%%%%%%%%%%%%%%%%%%%%%%%%%%%%%%%%%%%%%%%%%%
Consider a system of $N$ point-particles which interact via electrostatic forces. Each particle is specified by its mass, position $\vec{r}_i$ and charge $q_i$. The particles are contained in a cubic box $\Omega$ of length $L$ with periodic boundary conditions. The Coulomb potential $\phi$ at position $\vec{r}$ can be obtained by solving the equation
\begin{equation}
  -\Delta \phi(\vec{r}) = 4\pi \rho(\vec{r}),\qquad\text{with\quad $\rho(\vec{r})=\sum_{\vec{\mathfrak{n}}\in \mathbb{Z}^3} \sum_{j=1}^N q_j \delta(\vec{r}-\vec{r}_j-L\vec{\mathfrak{n}})$}
\label{eqn:Poisson}
\end{equation}
where $\rho(\vec{r})$ is the charge distribution of the particles. The first sum over $\vec{\mathfrak{n}}$ extends over all periodic copies of the box. 
The total long range potential of particle $i$ is
\begin{equation*}
  U^{(\lr)}(\vec{r}_i;\{\vec{r}_j\})=\sum_{j\ne i} U^{(\lr)}(\vec{r}_i,\vec{r}_j) = q_i\phi(\vec{r}_i).
\end{equation*}
To calculate the potential of a single point-charge, we rewrite the $\delta$-function as
\begin{equation*}
\begin{aligned} 
  \delta(\vec{r}) &= D^{(\sr)}(\vec{r}) + D^{(\lr)}(\vec{r})~\text{with}~D^{(\sr)} = S_\alpha(\vec{r}),\; D^{(\lr)}(\vec{r}) =\delta(\vec{r}) - S_\alpha(\vec{r})
\end{aligned}
\end{equation*}
where $S_{\alpha}(\vec{r})$ is a function which only depends on the distance $|\vec{r}|$ from the origin, integrates to 1 and decays exponentially as $\vec{r}\rightarrow \infty$. The length scale $\alpha^{-1/2}$ characterises the speed of this decay. This split of the $\delta$-function induces a separation in the potential into a short- and a long-range part with $\phi = \phi^{(\sr)} +  \phi^{(\lr)}$,
\begin{equation*}
  -\Delta \phi^{(*)}(\vec{r}) = 4\pi \rho^{(*)}(\vec{r}) = 4\pi \sum_{\vec{\mathfrak{n}}\in\mathbb{Z}^3}\sum_{j=1}^N q_j D^{(*)}(\vec{r}-\vec{r}_j-L\vec{\mathfrak{n}})~\text{where $*\in\{\sr,\lr\}$}.
\end{equation*}
Due to the construction of $S_\alpha$ the resulting short-range potential $\phi^{(\sr)}$ decays exponentially. This part can be safely truncated (see Section \ref{sec:error_analysis} for error estimates) and calculated with a local \texttt{PairLoop}  with a cutoff $r_c\gg \alpha^{-1/2}$. The long-range potential $\phi^{(\lr)}$ on the other hand is calculated in Fourier space.
%%%%%%%%%%%%%%%%%%%%%%%%%%%%%%%%%%%%%%%%%%%%%%%%%%%%%%%%%%%%%%%%%%%%%
\subsubsection{Short range potential}
%%%%%%%%%%%%%%%%%%%%%%%%%%%%%%%%%%%%%%%%%%%%%%%%%%%%%%%%%%%%%%%%%%%%%
To calculate the short-range potential, let $S_\alpha$ be a Gaussian with width $(2\alpha)^{-1/2}$
\begin{equation*}
  S_\alpha(\vec{r}) = \left(\frac{\alpha}{\pi}\right)^{\frac{3}{2}}\exp\left(-\alpha \vec{r}^2\right).
\end{equation*}
Then the short-range potential is readily evaluated as 
\begin{equation*}
  \phi^{(\sr)}(\vec{r}) = \sum_{\vec{\mathfrak{n}}\in\mathbb{Z}^3}\sum_{j=1}^N q_j \frac{\operatorname{erfc}\left(\sqrt{\alpha}|\vec{r}-\vec{r}_j-L\vec{\mathfrak{n}}|\right)}{|\vec{r}-\vec{r}_j-L\vec{\mathfrak{n}}|},\quad
\operatorname{erfc}(x) = \frac{2}{\sqrt{\pi}}\int_x^\infty e^{-t^2}\;dt.
\end{equation*}
%where $\operatorname{erfc(x)}$ is the complementary error function.
%%%%%%%%%%%%%%%%%%%%%%%%%%%%%%%%%%%%%%%%%%%%%%%%%%%%%%%%%%%%%%%%%%%%%
\subsubsection{Long range potential}
%%%%%%%%%%%%%%%%%%%%%%%%%%%%%%%%%%%%%%%%%%%%%%%%%%%%%%%%%%%%%%%%%%%%%
The long-range potential is evaluated in Fourier space. The Fourier-space representation of the charge distribution $\rho^{(\lr)}$ is given by
\begin{equation*}
  \hat{\rho}^{(\lr)}(\vec{k}) = \int_\Omega e^{-i\vec{k}\cdot\vec{r}} \rho^{(\lr)}(\vec{r})\; d\vec{r} = \sum_{j=1}^N q_j e^{-i\vec{k}\cdot\vec{r}_j}\exp\left(-\frac{\vec{k}^2}{4\alpha}\right).
\end{equation*}
The periodic boundary conditions restrict possible values of the reciprocal vector to $\vec{k}=(2\pi/L)\mathfrak{m}$ where $\mathfrak{m}\in\mathbb{Z}^3$. Since the Fourier-space representation of the Poisson equation in Eq. (\ref{eqn:Poisson}) is diagonal and given by $\vec{k}^2 \hat{\phi}(\vec{k}) = 4\pi \hat{\rho}(\vec{k})$, the long range potential in real space can be calculated as\footnote{Since we consider only neutral systems the term $\vec{k}=0$ can be dropped.}
\begin{equation*}
  \phi^{(\lr)}(\vec{r}) = \frac{1}{V}\sum_{\vec{k}} e^{i\vec{k}\cdot\vec{r}}\hat{\phi}(\vec{k}) = \frac{1}{V}\sum_{\vec{k}\ne 0}\sum_{j=1}^N\frac{4\pi}{\vec{k}^2}q_j e^{i\vec{k}\cdot(\vec{r}-\vec{r}_j)}\exp\left(-\frac{\vec{k}^2}{4\alpha}\right).
\end{equation*}
The second factor decays exponentially and the sum over $\vec{k}$ can be truncated for all $\vec{k}$ with $|\vec{k}| > k_c\gg \alpha^{1/2}$.

%%%%%%%%%%%%%%%%%%%%%%%%%%%%%%%%%%%%%%%%%%%%%%%%%%%%%%%%%%%%%%%%%%%%%
\subsubsection{Error estimate and computational complexity}\label{sec:error_analysis}
%%%%%%%%%%%%%%%%%%%%%%%%%%%%%%%%%%%%%%%%%%%%%%%%%%%%%%%%%%%%%%%%%%%%%
The short-range cutoff $r_c$ and long range-cutoff $k_c$ have to be carefully balanced to minimise the total error. As expected from dimensional analysis and discussed in detail in \cite{Kolafa1992} (see also \cite[Chapter 12]{Frenkel2001}), the error in both contributions to the potential is equal if $r_c \propto\alpha^{-1/2}$ and $k_c \propto\alpha^{1/2}$. For fixed density $\rho=N/V$ the number of particle-pairs considered in the short-range kernel is $4\pi/3\rho N r_c^3$ and the total cost for this part of the calculation is $\tau^{(\sr)}\propto N^2/V \alpha^{-3/2}$. Similarly the number of Fourier-modes in the long range calculation is $1/(6\pi^2)Vk_c^3$, and hence the total time in the long range part is $\tau^{(\lr)}\propto NV\alpha^{3/2}$. The optimal $\alpha$ which minimises the total calculation time $\tau=\tau^{(\sr)}+\tau^{(\lr)}$ is given by $\alpha\propto(N/V^2)^{1/3}$, which results in a computational complexity of $\tau=\mathcal{O}(N^{3/2})$. This also implies that the number $N_k$ of Fourier-modes in the long range calculation is $N_k\propto N^{1/2}$. 

The evaluation of the long-range potential at position $\vec{r}_i$ can be written as
\begin{equation}
  \phi^{(\lr)}(\vec{r}_i) = \sum_{\vec{k}\ne0}^{|\vec{k}|<k_c} C_{\vec{k}} A_{i,\vec{k}} \sum_{j=1}^N A^*_{j,\vec{k}} q_j
=
\sum_{\vec{k}\ne0}^{|\vec{k}|<k_c} C_{\vec{k}} A_{i,\vec{k}} \hat{\rho}_{\vec{k}},\qquad
\hat{\rho}_{\vec{k}} = \sum_{j=1}^N A^*_{j,\vec{k}} q_j\label{eqn:perf_mod1}
\end{equation}
with $A_{j,\vec{k}}:=\exp\left(i\vec{k}\cdot \vec{r}_j\right)$ and $C_{\vec{k}}:=4\pi/(V\vec{k}^2)\exp\left(-\vec{k}^2/(4\alpha)\right)$. The expression in Eq. (\ref{eqn:perf_mod1}) is essentially the product of a $N_k\times N$ matrix with a vector of length $N$ followed by a multiplication by an $N\times N_k$ matrix. Since the particles are distributed between the processors, but all Fourier modes computed on each processor, the computational cost is $\propto N N_k/p \propto N^{3/2}/p$. Every processor only calculates the contribution of all \textit{locally} stored particles to every Fourier mode. Combining the contributions of all particles to each of the $N_k$ Fourier modes therefore requires a global reduction of $N_k\propto N^{1/2}$ numbers, resulting in a total computational cost of $t = CN\left(\frac{N^{1/2}}{p} + r N^{-1/2}\log p\right)$ where the ratio $r\gg 1$ depends on the relative cost of computation and communication on a particular machine. We expect the code to scale well as long as $N\gg rp\log p$.
%%%%%%%%%%%%%%%%%%%%%%%%%%%%%%%%%%%%%%%%%%%%%%%%%%%%%%%%%%%%%%%%%%%%%
\subsection{Implementation}\label{sec:implementation}
%%%%%%%%%%%%%%%%%%%%%%%%%%%%%%%%%%%%%%%%%%%%%%%%%%%%%%%%%%%%%%%%%%%%%
%--------------------------------------------------------------------
\subsubsection{Short range potential}
%--------------------------------------------------------------------
By construction the short-range potential $\phi^{(\sr)}(\vec{r})$ rapidly converges to zero as the inter-particle distance $|\vec{r}|$ increases.
We truncate the short-range contribution to the electrostatic potential and force with a cutoff $r_c$ (see Section \ref{sec:error_analysis}),
\begin{align}
\phi^{(\sr)}_{r_c}(\vec{r}) &= \sum_{\substack{j~\text{with} \\ |\vec{r} - \vec{r}_j |<r_c}} q_j \frac{\operatorname{erfc}\left(\sqrt{\alpha}|\vec{r}-\vec{r}_j|\right)}{|\vec{r}-\vec{r}_j|}
\label{eqn:short_range_force}\\
\vec{F}^{(\sr)}_{r_c}(\vec{r}) &= \sum_{
\substack{j~\text{with} \\ |\vec{r} - \vec{r}_j |<r_c}}q_i q_j\frac{\vec{r}-\vec{r}_j}{|\vec{r}-\vec{r}_j|^2} \bigg[ \frac{\operatorname{erfc}\left(\sqrt{\alpha}|\vec{r}-\vec{r}_j|\right)}{|\vec{r}-\vec{r}_j|} + 2\sqrt{\frac{\alpha}{\pi}} \exp(-\alpha |\vec{r}-\vec{r}_j|^2)\bigg].\nonumber
\end{align}
The computational kernel for the \textit{local} \texttt{ParticlePair} loop is given in Listing \ref{lst_sr}. The position and charge data are stored per particle in \texttt{ParticleDat} data objects. Similarly, the resulting forces and total potential energy are stored as a \texttt{ParticleDat} and a \texttt{GlobalArray} object. Listing \ref{lst_pysr} shows the corresponding Python code for launching the pair loop. In the C-kernel capitalised variables such as \verb!REAL_CUTOFF_SQ! are constants which are replaced by their numerical values at compile time using the \verb!kernel_consts! dictionary.
\begin{lstlisting}[language={{c}},caption={Implementation of the short range force in Eq. (\ref{eqn:short_range_force}) and total electrostatic energy in the DSL for a \textit{Local Particle Pair Loop}. Output: short-range potential energy $u^{(\sr)}=\sum_{i=1}^{N}U^{(\sr)}_i$, $U^{(\sr)}_i = q_i \phi^{(\sr)}_{r_c}(\vec{r}_i)$ and short-range forces $\vec{F}^{(\sr)}_{r_c}(\vec{r}_i)$.}, label=lst_sr]
double r0 = r.j[0] - r.i[0];
double r1 = r.j[1] - r.i[1];
double r2 = r.j[2] - r.i[2];
double r_sq = r0*r0 + r1*r1 + r2*r2; double r = sqrt(r_sq);
double mask = (r_sq < REAL_CUTOFF_SQ)? 1.0 : 0.0;
double r_m1 = 1.0/r;
double qiqj_rm1 = q.i[0] * q.j[0] * r_m1 * mask;
double term1 = qiqj_rm1*erfc(SQRT_ALPHA*r);
u[0] += 0.5*term1; // electrostatic energy
double term3 = -1.*r_m1*(qiqj_rm1 * TWO_SQRT_ALPHAOPI * exp(MALPHA*r_sq) + r_m1*r_m1*term1); // force
F.i[0] += term3 * r0; F.i[1] += term3 * r1; F.i[2] += term3 * r2;
\end{lstlisting}
\mbox{%
\begin{lstlisting}[language={[ppmd]{python}},caption={
    Python \textit{local} \texttt{ParticlePair} loop creation and execution that reads \texttt{ParticleDat}s for positions $\vec{r}_i$ and charges $q_i$ and increments the \texttt{ParticleDat} for the force $\vec{F}_{r_c}^{(\sr)}$ and \texttt{GlobalArray} $u^{(\sr)}$.
}, label=lst_pysr]
# Define kernel
kernel = Kernel('ewald_sr', kernel_code, kernel_consts)
# Define and execute pair loop
pair_loop = PairLoop(kernel=kernel, shell_cutoff=rc,
                     dat_dict={'r': Positions(access.READ),
                               'q': Charges(access.READ),
                               'F': Forces(access.INC),
                               'u': u_sr(access.INC)})
pair_loop.execute()
\end{lstlisting}
}
%--------------------------------------------------------------------
\subsubsection{Long range potential}
%--------------------------------------------------------------------
The computation of the long-range potential is split into two \texttt{ParticleLoop}s which correspond to the $N_k\times N$ and $N\times N_k$ matrix-vector products described in Section \ref{sec:error_analysis}. The first iterates over all particles $j$ and for each particle computes the contribution to $\hat{\rho}_{\vec{k}}$ defined in Eq. (\ref{eqn:perf_mod1}) for all $|\vec{k}|<k_c$. An outline of the computational kernel is shown in Algorithm \ref{alg:lr1} (for brevity we do not show the corresponding C- and Python-code, but outline the access descriptors). We order the entries in the \texttt{GlobalArray} $\hat{\rho}_{\vec{k}}$ such that loops over reciprocal vectors $\vec{k}$ are vectorised by the compiler (as confirmed by the generated assembly code).
\begin{algorithm}[H]
\caption{Computational kernel for \texttt{ParticleLoop} I.\newline \textit{Input}: position $\vec{r}_j$ [\texttt{READ}], charge $q_j$ [\texttt{READ}]. \textit{Output}: reciprocal space $\hat{\rho}_{\vec{k}}$ [\texttt{INC}] }
\label{alg:lr1}
\begin{center}
\begin{algorithmic}[1]
\FORALL{reciprocal vectors $\vec{k}\ne 0$ such that $|\vec{k}|<k_c$}
  \STATE{$\hat{\rho}_{\vec{k}} \mapsto \hat{\rho}_{\vec{k}} +  A^*_{j,\vec{k}} q_j$}
\ENDFOR
\end{algorithmic}
\end{center}
\end{algorithm}
Note that the calculation of $\hat{\rho}_{\vec{k}}$ requires global reductions since each $\vec{k}$-component receives contributions from all particles in the system. This, however, is automatically handled by the code generation system and requires no explicit coding for the user who only writes the local kernel in line 2 of Algorithm \ref{alg:lr1}. In our implementation we store copies of the entire vector $\hat{\rho}_{\vec{k}}$ on each MPI task and do not attempt a parallel domain decomposition in $\vec{k}$ space. Since the number of reciprocal vectors grows $\propto{\sqrt{N}}$ this does not lead to memory issues for moderately sized systems for which the Particle-Ewald method is competitive.

Given the vector $\hat{\rho}_{\vec{k}}$, the electrostatic energies and forces are calculated as a second \texttt{ParticleLoop} using Eq. (\ref{eqn:perf_mod1}) for each particle in Algorithm \ref{alg:lr2}.
\begin{algorithm}%[H]
    \caption{Computational kernel for \texttt{ParticleLoop} II.\newline \textit{Input}: position $\vec{r}_j$ [\texttt{READ}], charge $q_j$ [\texttt{READ}], $\hat{\rho}_{\vec{k}}$ [\texttt{READ}]. \textit{Output}: total electrostatic potential energy $u^{(\lr)}$ [\texttt{INC}] and forces $\vec{F}^{(\lr)}_j\equiv \vec{F}^{(\lr)}(\vec{r}_j)$ [INC]. }
\label{alg:lr2}
\begin{center}
\begin{algorithmic}[1]
\FORALL{reciprocal vectors $\vec{k}\ne 0$ such that $|\vec{k}|<k_c$}
  \STATE{$u^{(\lr)} \mapsto u^{(\lr)} +  C_k A_{j,\vec{k}} q_j \hat{\rho}_{\vec{k}}$}
  \STATE{$\vec{F}^{(\lr)}_j \mapsto \vec{F}^{(\lr)}_j - i\vec{k} C_k A_{j,\vec{k}} q_j \hat{\rho}_{\vec{k}}$}
\ENDFOR
\end{algorithmic}
\end{center}
\end{algorithm}

The self-energy (not shown here) is calculated once at the beginning of the simulation and the cost of this operation is amortised over the total runtime.
%--------------------------------------------------------------------
\subsubsection{Hybrid parallelisation with OpenMP}
%--------------------------------------------------------------------
In MPI-only mode, the simulation domain $\Omega$ is split into local sub-domains which are distributed across CPU cores as described in \cite{ppmd2017}. To extend and improve scalability, we also implemented an MPI+OpenMP hybrid mode. In this case each node (MPI-rank) handles one sub-domain and the particles in this local domain are distributed across OpenMP threads. As discussed in \cite{ppmd2017} we require that \texttt{ParticleLoop}s only write to one particle and therefore no special approaches to handle write conflicts such as colouring are required in this case. To deal with potential write conflicts in \texttt{GlobalArray} operations, thread safe reduction code is generated outside the C-kernel written by the user.
%%%%%%%%%%%%%%%%%%%%%%%%%%%%%%%%%%%%%%%%%%%%%%%%%%%%%%%%%%%%%%%%%%%%%
\section{Results}\label{sec:results}
%%%%%%%%%%%%%%%%%%%%%%%%%%%%%%%%%%%%%%%%%%%%%%%%%%%%%%%%%%%%%%%%%%%%%
%%%%%%%%%%%%%%%%%%%%%%%%%%%%%%%%%%%%%%%%%%%%%%%%%%%%%%%%%%%%%%%%%%%%%
\subsection{Computational Complexity} \label{sec:results_order}
\begin{figure}%[H]
\ifbool{PREPRINT}{%
\begin{center}
}{}
\includegraphics[width=.55\linewidth]{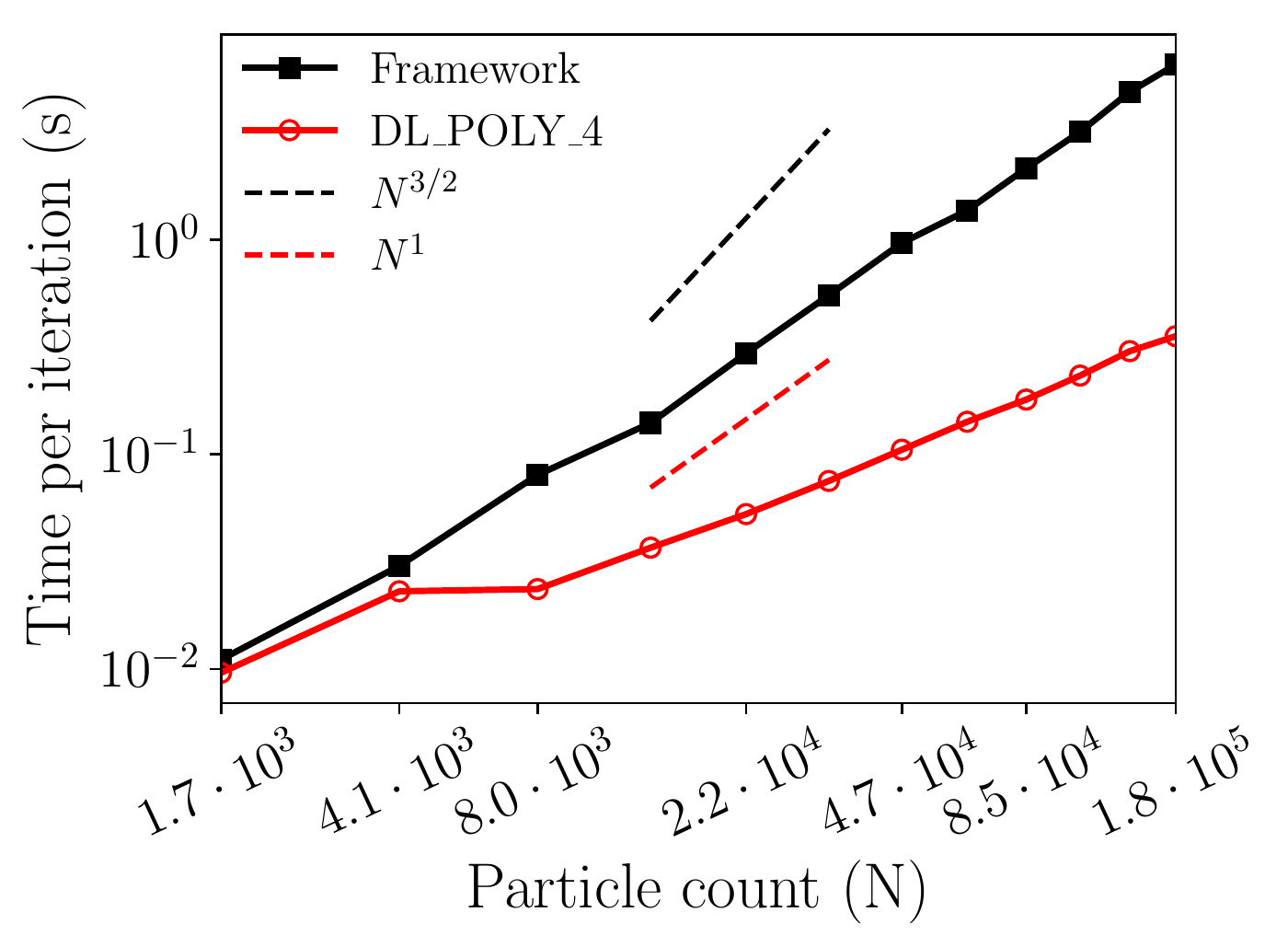}
\caption{Time per iteration against particle count for an NaCl system on a single 8 core CPU using OpenMP (our framework) or pure MPI (DL\_POLY\_4).}
\label{fig:order}
\ifbool{PREPRINT}{%
\end{center}
}{}
\end{figure}

%%%%%%%%%%%%%%%%%%%%%%%%%%%%%%%%%%%%%%%%%%%%%%%%%%%%%%%%%%%%%%%%%%%%%
With correct choice of $\alpha$ the Ewald method exhibits $\mathcal{O}(N^{3/2})$ computational cost. Figure \ref{fig:order} confirms this by plotting the time per iteration for a NaCl salt simulation against particle count $N$ at a fixed density of 1 atom per (2.5\AA)$^3$. We include repulsive Lennard-Jones interactions to prevent the particle distribution from collapsing. However, for sizable particle counts the dominant computational cost are electrostatic forces: for $N=1.8\cdot 10^5$ particles 87\% of the time is spent computing Coulombic interactions. For all tests in this paper we set the error tolerance to $10^{-6}$ and vary the parameters $\alpha$ and $r_c$ (which balance the work between the real- and Fourier-space) to minimise the runtime. For our framework the pair $(\alpha,r_c)$ takes values between $(0.062,13.5\text{\AA})$ for $N=1728$ and $(0.013,29.2\text{\AA})$ for $N=1.8\cdot10^{5}$. For DL\_POLY\_4 \cite{Todorov2006} we choose a cutoff value of $r_c=10\text{\AA}$. All runs are carried out on the ``Balena'' cluster; one node consists of two Intel E5-2650v2 8-core CPUs.
\noindent
Both implementations show better than expected scaling with $N$. For small particle numbers the SPME method used by DL\_POLY\_4 is in the same ballpark as our implementation. The SPME method obviously outperforms our method for larger particle counts where it is an order of magnitude faster.
\subsection{Strong Scaling} \label{sec:results_scaling}
To study the parallel scalability we set the number of particles to $N=3.3\cdot10^{4}$ in a box of size $80\text{\AA}\times80\text{\AA}\times80\text{\AA}$ (at the same density as in Section \ref{sec:results_order}) and increase the core count. The spatial domain cannot be decomposed into regions of side length less than the cutoff $r_c$ which prevents repeating the runs in Section \ref{sec:results_order} on more than one node. To address this, we fixed $r_c=19\text{\AA}$ ($r_c=10\text{\AA}$ for DL\_POLY\_4) at the price of using a non-optimal value of $\alpha$ ($0.032$ instead of $0.023$). This allows to extend the scalability of the MPI-only implementation and DL\_POLY\_4 to 64 cores and we find that it has no negative impact on the runtime on one CPU. To scale beyond this limit we use the hybrid MPI+OpenMP scheme with one MPI process per CPU socket to run on up to 256 cores. To quantify any potential performance loss due to the non-optimal value of $\alpha$, we also include the relevant data point with $(\alpha,r_c)=(0.023,22.1\text{\AA})$ from Fig. \ref{fig:order}.
\begin{figure}%[H]
    \begin{subfigure}[t]{.48\textwidth}
        \includegraphics[width=1.0\linewidth]{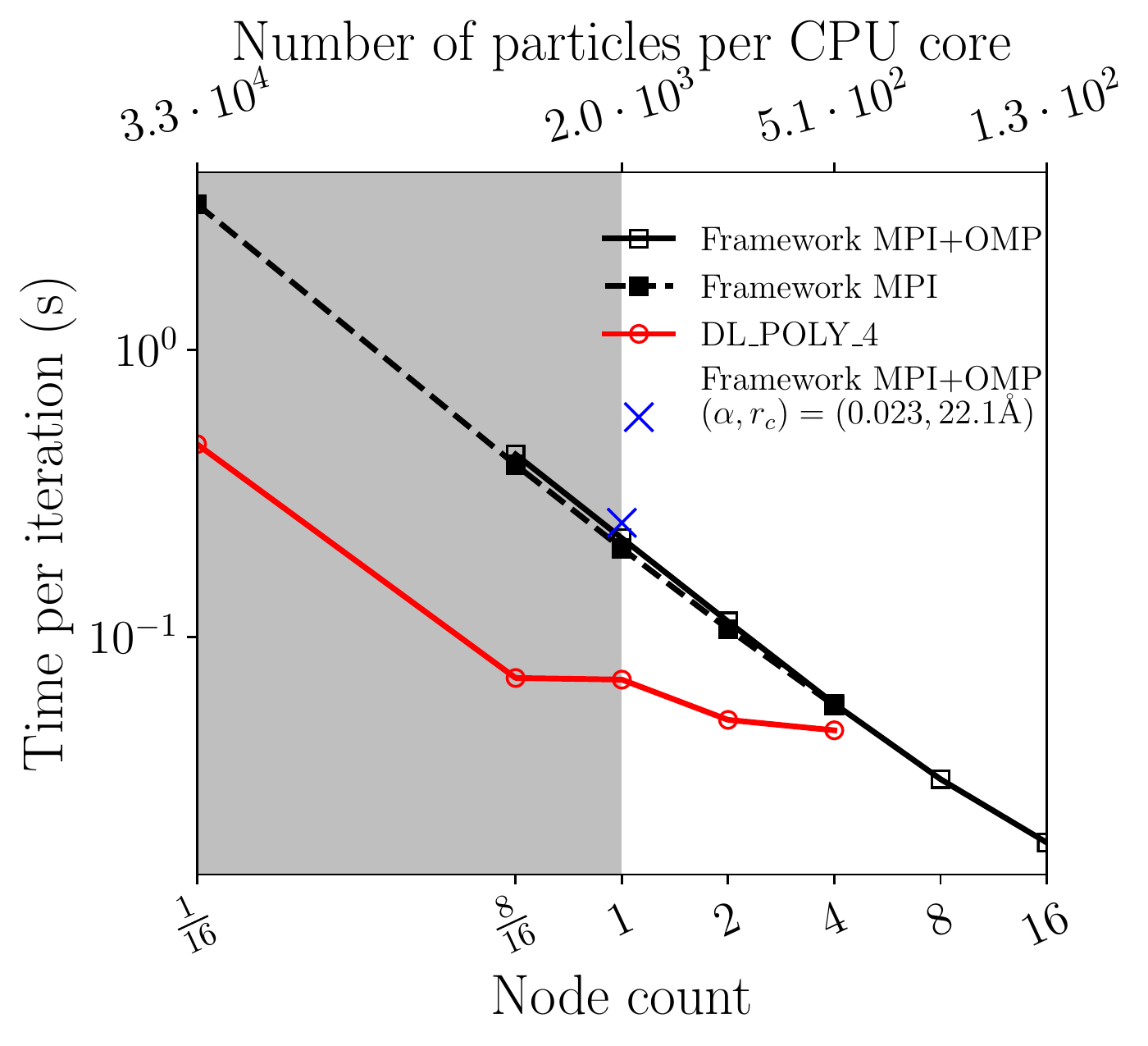}
    \end{subfigure}
    ~
    \begin{subfigure}[t]{.48\textwidth}
        \includegraphics[width=1.0\linewidth]{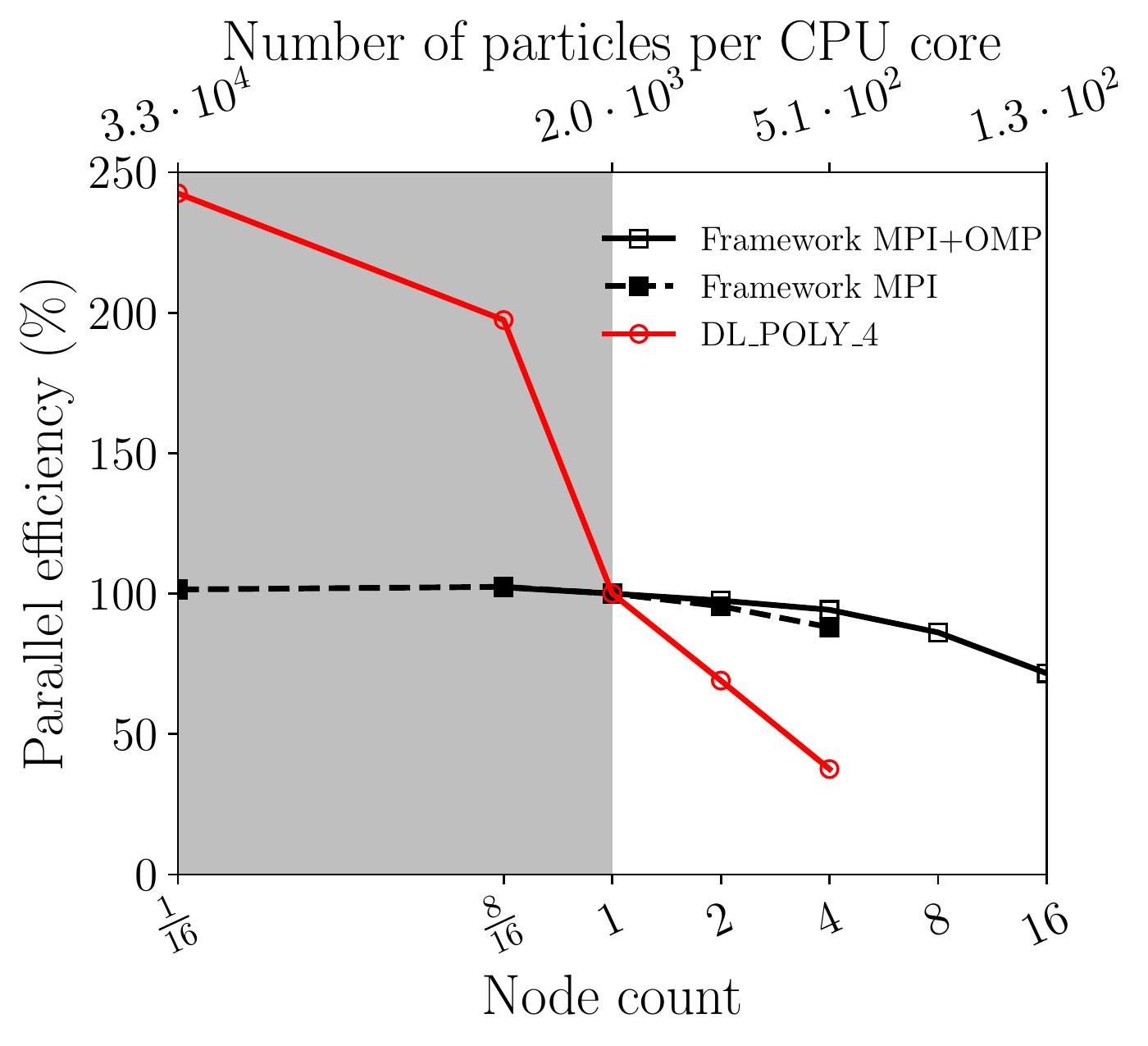}
    \end{subfigure}

    \caption{Strong scaling experiment of an NaCl system comparing the framework with DL\_POLY\_4. Time per iteration (left) and parallel efficiency relative to one 16-core node (right).}
    \label{fig:strong_time}
\end{figure}
Both the MPI and MPI+OpenMP implementations exhibit decent scaling to 16 nodes (256 cores). DL\_POLY\_4 is faster overall on smaller core counts but does not scale to larger core counts. The MPI+OpenMP execution of Algorithm \ref{alg:lr2} on one node achieved an average of 34\% of peak floating point vector performance.
The computationally most expensive component is the loop over all Fourier modes $\vec{k}=(k_1,k_2,k_3)$. This has been vectorised over the four quadrants with $(\operatorname{sign}(k_1),\operatorname{sign}(k_2))=(+,+)$, $(+,-)$, $(-,+)$ and $(-,-)$ and we confirmed that the Intel compiler indeed generates packed vector instructions.
%%%%%%%%%%%%%%%%%%%%%%%%%%%%%%%%%%%%%%%%%%%%%%%%%%%%%%%%%%%%%%%%%%%%%
\section{Conclusion}\label{sec:conclusion}
%%%%%%%%%%%%%%%%%%%%%%%%%%%%%%%%%%%%%%%%%%%%%%%%%%%%%%%%%%%%%%%%%%%%%
We demonstrated how the abstraction and Python code-generation system in \cite{ppmd2017} can be used to implement long-range electrostatic interactions in Molecular Dynamics simulations. Our Particle-Ewald implementation achieves good absolute performance and parallel scalability. In addition to \cite{ppmd2017} we now also support a hybrid MPI+OpenMP backend which is used to extend scalability in the strong scaling limit. To include long range forces for significantly larger systems we will investigate the implementation of SPME \cite{spme2} algorithms or the Fast Multipole Method (FMM) \cite{fmm1}. This will require adding new data structures such as a hierarchical meshes for FMM or linking to existing Fast Fourier Transform libraries.
%%%%%%%%%%%%%%%%%%%%%%%%%%%%%%%%%%%%%%%%%%%%%%%%%%%%%%%%%%%%%%%%%%%%%
\subsection*{Acknowledgements}
%%%%%%%%%%%%%%%%%%%%%%%%%%%%%%%%%%%%%%%%%%%%%%%%%%%%%%%%%%%%%%%%%%%%%
The PhD project of William Saunders is funded by an EPSRC studentship. This research made use of the Balena HPC service at the University of Bath.
\bibliographystyle{unsrt}
\ifbool{PREPRINT}{%
\apptocmd{\thebibliography}{\setlength{\itemsep}{-0.35ex}}{}{}
}{}
%\bibliography{paper}

\begin{thebibliography}{10}

\bibitem{Bertolli2012}
C.~Bertolli et~al.
\newblock {\em {in Euro-Par 2011}}, pages 191--200.
\newblock Springer, Berlin, Heidelberg, 2012.

\bibitem{Rathgeber2012}
F.~Rathgeber et~al.
\newblock In {\em HPC, Networking Storage and Analysis, SC Companion:}, pages
  1116--1123, Los Alamitos, CA, USA, 2012. IEEE Computer Society.

\bibitem{ppmd2017}
W.~R. Saunders, J.~Grant, and E.~H. Mueller.
\newblock {A Domain Specific Language for Performance Portable Molecular
  Dynamics Algorithms}.
\newblock {\em submitted to Computer Physics Communications, preprint:
  \href{https:/arxiv.org/abs/1704.03329}{arxiv:1704.03329}}, 2017.

\bibitem{Plimpton1995}
S.~Plimpton.
\newblock {\em Journal of Computational Physics}, 117(1):1 -- 19, 1995.

\bibitem{Todorov2006}
I.~T.~Todorov et~al.
\newblock {\em J. Mater. Chem.}, 16:1911--1918, 2006.

\bibitem{rapaport_art}
D.~C. Rapaport.
\newblock {\em The Art of Molecular Dynamics Simulation}.
\newblock Cambridge University Press, New York, 2nd edition, 2004.

\bibitem{Ewald1921}
P.~P. Ewald.
\newblock {\em Annalen der Physik}, 369(3):253--287, 1921.

\bibitem{spme2}
T.~Darden, D.~York, and L.~Pedersen.
\newblock {\em J. of Chem. Phys.}, 98(12):10089--10092, 1993.

\bibitem{spme3}
U.~Essmann et~al.
\newblock {\em J. of Chem. Phys.}, 103(19):8577--8593, 1995.

\bibitem{fmm1}
L.~Greengard and V.~Rokhlin.
\newblock {\em Journal of Comp. Phys.}, 73(2):325--348, 1987.

\bibitem{Kolafa1992}
J.~Kolafa and J.~W. Perram.
\newblock {\em Molecular Simulation}, 9(5):351--368, 1992.

\bibitem{Frenkel2001}
D.~Frenkel and B.~Smit.
\newblock {\em {Understanding molecular simulation}}, volume~1.
\newblock Academic Press, 2001.

\end{thebibliography}

\end{document}